
\magnification=\magstep1
\baselineskip=24 true pt
\hsize=33 pc
\vsize=45 pc
\rightline {IP/BBSR/92-38}
\rightline {May, 92}
\def \clg {{\cal G}}
\bigskip
\centerline {\bf SYMMETRIES OF STRING EFFECTIVE ACTION}
\centerline {\bf AND SPACE-TIME GEOMETRY}
\bigskip
\bigskip
\bigskip
\centerline {{\bf S. Pratik Khastgir}$^{\star}$ }
\smallskip
\centerline {\it { Institute of Physics, Bhubaneswar-751005, INDIA}}
\bigskip
\centerline { and}
\bigskip
\centerline {{\bf Jnanadeva Maharana}$^{\dagger}$}
\smallskip
\centerline {\it { California Institute of Technology, Pasadena, CA 91125}}
\bigskip
\bigskip
\bigskip
\centerline {\bf Abstract}

Two dimensional charged black hole solution is obtained by implementing
an $O(2,2)$ transformation on the three dimensional black string solution.
Two different monopole backgrounds in five dimensions are related
through an $O(2,2)$ transformation. It has been shown in these examples that
the particular $O(2,2)$ transformation corresponds to duality transformation.

\bigskip
\bigskip
\bigskip
\bigskip
\bigskip
\bigskip
\hrule

$\star$ e-mail: pratik$\%$iopb@shakti.ernet.in

$\dagger$ Permanent address:Institute of Physics, Bhubaneswar-751005, INDIA

$\;\;\;$e-mail: maharana@theory3.caltech.edu until June 30.

\vfil
\eject
\noindent {\bf I. Introduction}

  The string theory offers the prospect of unifying all the four fundamental
forces of nature. In the first quantized framework, one considers the
evolution of the string in the background of its massless modes. This
approach has proved to be quite powerful in undrstanding various aspects of
the string theory. The requirement of conformal invarience of string theory
imposes stringient constraints on the configurations of the background fields.
In other words, if we require the beta-functions associated with these
backgrounds to vanish, then we arrive at differential equations known as the
equations motion for the background fields. It can be argued that so long as
we are interested in the low energy effects of the string theory, it suffices
to take into account the effects of the massless modes of the string.
We can construct the tree level string effective action involving only the
massless excitations in such a way that the equations of motion
derived from the effective action exactly correspond to the
requirements of the vanishing of the beta-functions [1] as discussed above.

  Recently, there has been an extensive study of the properties of the string
effective action in order to explore the consequences of string theory in
cosmology [2] as well as to investigate properties of black holes [3].
Other interesting types of space-time structures like strings and branes
[4,5] have also emerged in this investigation. The monopole and dyon [6,7]
solutions were also obtained in this context. Furthermore many of these
solutions can be shown to correspond to exact conformal field theory [5,8,9].
In this context,
a rich symmetry structure of the string effective action has been unravelled.
Indeed, these symmetry properties have played an important role in discovering
new vacuum configurations of the string theory ( in general they correspond
to inequivalent string vacua ). These vacuum configurations are identified as
different spacetime geometries.

 We may recall that target space duality is a very important symmetry
property of the string theory. A familiar example is the $R$ duality [10]: when
we
consider a compactified string on a circle of radius $R$, the spectrum of this
string remains invariant under the transformation $R \rightarrow
{\lambda_{s}^{2}/ R}$
($\lambda_{s}^{2} = 2\alpha'\hbar$). The consequences of duality are
far reaching than
that meets the eye and have been explored in several directions. It has led to
introduction of a minimum compactification scale. The idea of duality has been
employed to restrict the form of scalar (or super) potentials and to study
nonperturbative supersymmetry breaking. Furthermore, the consequences of
duality
in cosmology [11] have turned out to be quite interesting and imoprtant.

   Another important symmetry property of the string efective action has
been discovered by Meissner and Veneziano [12] when the massless
backgrounds are
only allowed to depend on time. It was shown that that the effective action is
invaraint under global $O(d,d)$ transformations where $d$ is the spatial
dimension and
spacetime dimension, $D = d + 1$. Subsequently, it was shown that one can
generate
new cosmological solutions [13] from a given solution. In particular, it was
demonstrated that it was possible to generate spacetime with nontrivial
geometries by implementing $O(d,d)$ transformations on an initial background
with trivial (flat geometries [13]). Furthermore, a large class of black
hole solutions have been generated through suitable choice of $O(d,d)$
transformations [14].

    The purpose of this article is to explore further the consequences of
the $O(d,d)$ transformations in string theory. We demonstrate that
given a background space-time geometry which satisfies vanishing beta-function
constraints, it is possible to generate a new class of background
configurations through $O(d,d)$
transformations. It is shown that a two dimensional charged
black hole solution [9] can be obtained by implementing a suitable $O(d,d)$
transformation on the three dimensional
black string solution of Horne-Horowitz [5]. Furthermore, we
explicitly demostrate that starting from a generalized
Kaluza-Klein monopole solution obtained by Sorkin and Gross and Perry [6]
and studied  by Banks et al. [15], in the context of string theory, another
solution  of Ref. [15] can be generated by an appropiate
$O(d,d)$ transformation.
Moreover, we are able to provide an interpretation of these
transformations from the duality point of view discussed in Refs. [16,17].

   The paper is organized as follows: In section II we recall some of the
important results of Meissner and Veneziano in the context of $O(d,d)$
symmetries of string effective action and discuss how to generate new
solution. Then we describe briefly isometries of target space backgruond
fields and their relation to duality transformations.
Section III deals with the specific string theoretic models we are interested
in and the configurations of the background fields.
We introduce explicit $O(d,d)$
transformations and obtain the new background field configurations. The
sigma model interpretation of the newly generated backgrounds is discussed.
Next, we elucidate on the relations of our work with the results of Ref
[16,17]. The summary and conclusions are given in Sec IV.

\noindent {\bf II. The symmetries of string effective action}

In this section we shall try to summarize some of the important results
obtained by Veneziano, Meissner, and Sen [12,14]. We shall also discuss
the duality transformations briefly [16,17] at the end.

We consider a closed bosonic string in the background of its massless
modes such as graviton, dilaton and antisymmetric tensor fields. The
equations of motion of the tree level string effective action given
below corresponds to the requirement of the vanishing of the
beta-functions associated with these background fields.
$$ S=-\int d^{D}x \sqrt{-detG}
e^{-\phi}\big [ R+G^{\mu\nu}\partial_{\mu}\phi\partial_{\nu}\phi+{1\over
12}H_{\mu\nu\rho}H^{\mu\nu\rho}-V\big ]\eqno(1)$$

\noindent where $V$ is the cosmological term proportional to $D-26$
(proportional to $D-10$ for superstring),
$\phi$ is the
dilaton field, $G_{\mu\nu}$ is the $D$-dimensional metric and $H_{\mu\nu\rho}$
is the field strength for the antisymmetric tensor field $B_{\mu\nu}$:

$$H_{\mu\nu\rho}=\partial_{\mu}B_{\nu\rho}+cyclic. \eqno(2)$$

If the backgrounds $G$, $B$, and $\phi$ are functions of only one coordinate
(say time $t$), then the metric $G$ and the antisymmetric tensor $B$
 can be written in the following form implementing the general coordinate
 transformations and the Abelian gauge transformations on $G$ and $B$
respectively.
$$G=\bigg (\matrix {{-1} & {0}\cr {0} &{ \clg (t)} \cr }\bigg ),
\qquad   B=\bigg (\matrix
{{0} & {0} \cr {0} & {{\cal B} (t)} \cr}\bigg ),\eqno(3)$$

\noindent  where $\clg (t)$ and ${\cal B}(t)$ are $d\times d$ matrices with
$d = D - 1$. Then the reduced action obtained from eqn. (1) can be written
in a manifestly $O(d,d)$ invariant form obtained by Meissner
and Veneziano [12].

$$ S=\int dt
e^{-\Phi}[V +\dot \Phi^{2}+{1\over 8}Tr({\dot M}\eta{\dot
 M}\eta)],\eqno(4)$$
\noindent with
$$\Phi=\phi-ln\sqrt{det \clg}, \eqno(5)$$

$$M \equiv\Bigg ( \matrix {{\clg^{-1}} & {-{\clg^{-1}{\cal B}} }\cr
{{\cal B}{\clg^{-1}}} &
{\clg-{\cal B}{\clg^{-1}}{\cal B}} \cr} \Bigg ),\eqno(6)$$

\noindent and
$$\eta =\Bigg ( \matrix {{0} & {I}\cr {I} & {0} \cr} \Bigg ).\eqno(7)$$
\noindent In last expression $I$ stands for $d$-dimensional unit matrix.

The action (4) is manifestly invariant under the global $O(d,d)$
group [12] acting as,
$$\Phi \rightarrow \Phi,\qquad  M \rightarrow \Omega M \Omega^{T}. \eqno (8)$$
\noindent Here $\Omega $ is an $O(d,d)$ matrix satisfying
$$\Omega \eta \Omega^{T}=\eta . \eqno(9)$$

This $O(d,d)$ transformations relate different string vacua (geometries)
which are not equivalent in general.
Recently, Sen [14] has shown in the frame work of string field theory
that the space of solutions has  $O(d)\times O(d)$ symmetry, which is
a subgroup of $O(d,d)$. Moreover, it is argued that the diagonal
subgroup $O(d)$ of $O(d)\times O(d)$ generates only spatial rotations. Thus,
the coset is equivalent to $(O(d)\times O(d))/O(d)$ and has
dimensionality $d(d-1)/2$ in contrast to the time independent case
where the dimensionality is $d^2$. It was shown [14] that $O(d)\times
O(d)$ symmetry persists to all orders in string tesion
$\alpha^{'}$. Recently, several interesting solutions (both cosmological
[13] and
black hole type [14]) were generated from known backgrounds exploiting this
symmetry.

Next we briefly describe the isometry and duality [16,17]. Starting
with graviton ($g_{\mu\nu}$), antisymmetric field tensor
($B_{\mu\nu}$), and dilaton ($\phi$) background, which has a
translational symmetry (isometry) in $x$, one can generate a dual background
given by [16],

$${\tilde g}_{xx}={1\over g_{xx}},\qquad  {\tilde g}_{ax}={B_{ax}\over
g_{xx}},\qquad {\tilde g}_{ab}={g_{ab}-{{g_{ax}g_{xb}+B_{ax}B_{xb}}\over
g_{xx}}}$$

$$ {\tilde B}_{ax}={g_{ax}\over
g_{xx}},\qquad {\tilde B}_{ab}={B_{ab}-{{g_{ax}B_{xb}+B_{ax}g_{xb}}\over
g_{xx}}},\qquad {\tilde \phi}= \phi-\ln g_{xx}, \eqno (10)$$
\noindent where $a$ and $b$ run over all directions except $x$.
The original background and its dual satisfy the same equations of motion.
This duality
of low energy field equations exists whether or not $x$
is compact. Recently it has been shown [17] that if $x$ is compact,
the original solution and dual are both low-energy approximation to
the same conformal field theory. This completes our recapitulation.

\noindent {\bf III. Applications of $O(d,d)$ transformation}

In this section we present two explicit examples where $O(d,d)$
transformations are implemented and dual solutions are generated. The
first example is in three dimensions and the other is in five dimensions.
The similarities of these two examples will be distinct when we describe the
examples.

Our starting point is a classical solution which was first
obtained by Horne and Horowitz [5] as an exact conformal field
theory. This result was obtained by gauging a one dimensional
subgroup of $G=SL(2,R)\times R$. In this consruction, which is a generalisation
of Witten's [8] construction, a free boson $x$ is added to the theory.
The action is,
\def \pr {\partial}
$$S={1\over \pi}\int_{\Sigma}d^{2}\sigma\Big [ {{k'\pr_+r\pr_-r}\over
{8r^2(1-{M\over r})(1-{Q^2\over Mr})}}-(1-{M\over
r})\pr_+t\pr_-t$$
$$+(1-{Q^2\over Mr})\pr_+x\pr_-x+{Q\over M}(1-{M\over
r})(\pr_+x\pr_-t-\pr_-x\pr_+t)\Big ]\eqno (11)$$
\noindent The space-time metric has the form,
$$ds^2 = -(1-{M\over r})dt^2 + (1-{Q^2\over {Mr}})dx^2 +
(1-{M\over r})^{-1}(1-{Q^2\over {Mr}})^{-1}{k'dr^2\over {8r^2}}\eqno(12)$$
\noindent whereas the antisymmetric tensor field and the dilaton are
given by,
$$B_{tx} = {Q\over M}(1-{M\over r}),\qquad \phi =- \ln r - {1\over 2}\ln
{k'\over 2}, \eqno (13) $$
\noindent $Q$ and $M$ are the axionic charge and mass per unit length
of the black string. The $k'$ is the WZW level. The central charge of
the theory is ${3k'\over {k'-2}}$. The equations of motion of the low
energy string effective action,

$$ S=\int d^{3}x \sqrt{detG}
e^{-\phi}\big [R+G^{\mu\nu}\partial_{\mu}\phi\partial_{\nu}\phi-{1\over
12}H^2 + {8\over k'}\big ].\eqno(14)$$
\noindent are the conditions for vanishing of $\beta$-functions. Here
${8\over k'}$ plays the role of cosmological constant.

We observe that the backgrounds are independent of two
coordinates $x$ and $t$. Thus one can perform an $O(2,2)$
transformation to generate a new solution and we take $\Omega \equiv O(2,2)$
(which is also an element of $O(2)\times O(2)$) as [14],
$$\Omega ={1\over 2}\Bigg ( \matrix {{S+R} & {R-S}\cr {R-S} & {S+R} \cr}
\Bigg ) .\eqno(15)$$
\noindent Note that $O(2)\times O(2)$ is a subgroup of $O(2,2)$ and
$S$ and $R$ are $O(2)$ matrices given by,

$$S=\Bigg ( \matrix {{1} & {0}\cr {0} & {-1} \cr}
\Bigg ),\qquad R=\Bigg ( \matrix {{1} & {0}\cr {0} & {1} \cr}
\Bigg ).\eqno(16)$$
\noindent We first identify the $x-t$ block of the metric and
antisymmetric tensor,

$$\clg = \Bigg ( \matrix {{-(1-{M\over r})} & {0}\cr {0} &
{(1-{Q^2\over {Mr}})} \cr}
\Bigg ),\qquad {\cal B} = \Bigg ( \matrix {{0} & {{Q\over M}(1-{M\over r})
}\cr {-{Q\over M}(1-{M\over r})} & {0} \cr}
\Bigg ).\eqno(17)$$
\noindent Now performing the $O(2,2)$ transformation on this block of
metric and antisymmetric tensor, we obtain new $ M' = \Omega^{T}M\Omega$ ,
$M$ defined earlier by eqn. (6).
The new backgrounds $\clg'$ and ${\cal B}'$ as,
$$\clg' = \Bigg ( \matrix {{-(1-{M\over r})+{{{Q^2\over
{M^2}}(1-{M\over r})^2} (1-{Q^2\over {Mr}})^{-1}}} & {
{{Q\over M}(1-{M\over r})}(1-{Q^2\over {Mr}})^{-1}}\cr
{{{Q\over M}(1-{M\over r})}(1-{Q^2\over {Mr}})^{-1}} &
{(1-{Q^2\over {Mr}})^{-1}} \cr}
\Bigg ), \qquad {\cal B}'=0. \eqno (18)$$
\noindent The metric in three dimensions takes the form,

$$ds^2 = \big [-(1-{M\over r})+{{{Q^2\over
{M^2}}(1-{M\over r})^2}(1-{Q^2\over {Mr}})^{-1}}\big ]dt^2 +
{{{2Q\over M}(1-{M\over r})}(1-{Q^2\over {Mr}})^{-1}}dtdx$$

$$+ {(1-{Q^2\over {Mr}})^{-1}}dx^2
+ (1-{M\over r})^{-1}(1-{Q^2\over {Mr}})^{-1}{k'dr^2\over {8r^2}},
\eqno (19)$$

\noindent and the new dilaton is given by (recall
$\Phi=\phi-\ln{\sqrt {det\clg}}$, expression (8), remains invariant
under this transformation),
$$\phi' = \phi - {1\over 2}\ln{det\clg \over {det\clg'}} =
 -\ln r({1-{Q^2\over{Mr}}}) -{1\over 2}\ln {k'\over 2}.  \eqno (20)$$
\noindent If we make the coordinate transformation
$$r \rightarrow {Q^2\over M} + M(1-{Q^2\over {M^2}})\cosh^2r, \eqno(21)$$
\noindent with

$${Q\over M} = e ,\qquad  {1\over 2}(1-{Q^2\over {M^2}}) = {k'\over
4}, \eqno (22)$$

\noindent the metric elements will reduce to,
$$g_{xx}={r\over {r-{Q^2\over M}}}\rightarrow {{{Q^2\over M}
+ M(1-{Q^2\over {M^2}})\cosh^2r}\over{M(1-{Q^2\over {M^2}})\cosh^2r}}=
1 +{2e^2\over
{k'}}{1\over {\cosh^2r}}\eqno(23a)$$
$$g_{xt}={Q\over M}{{(r-M)}\over {(r-{Q^2\over M})}}\rightarrow
{Q\over M}{{M(1-{Q^2\over {M^2}})(\cosh^2r-1)}\over
{M(1-{Q^2\over {M^2}})\cosh^2r}}=e\tanh^2r\eqno (23b)$$
$$g_{rr}\rightarrow {k'\over2},\quad etc..  \eqno(23c)$$
\noindent The metric in three dimensions and the dilaton read,

$$ds^2 = {k'\over2}dr^2-{k'\over2}\tanh^2rdt^2 + 2e\tanh^2rdxdt +
(1 +{2e^2\over {k'}}{1\over {\cosh^2r}})dx^2 \eqno(24a)$$
\noindent and $$\phi = -\ln\cosh^2r + const. \eqno (24b)$$
\noindent If we take $x$ as a compact direction this is exactly
the 2-dimensional charged black hole solution obtained originally by
Ishibashi, Lie and Steif [9] by gauging a subgroup $U(1)$ of $SU(2)\times
U_{i}(1)$, where $i$ denotes the internal direction. The
2-dimensional metric is,
$$ds^2 = 2kdr^2-2k\tanh^2rdt^2\eqno(25a)$$
\noindent and the gauge field,
$$A_r = 0 ,\quad A_t = \tanh^2r. \eqno(25b)$$
\noindent There are two scalar backgrounds such as dilaton and the $``$Higgs"
field given by,
$$\phi = -\ln \cosh^2r+const. ,\qquad  \psi = (1 +{e^2\over
{2k}}{1\over {\cosh^
2r}}). \eqno (25c)$$
\noindent We mention in passing that the parameter $k'$, appearing in
the WZW action given by eqn. (11) is related to the parameter of Ishibashi
$et al.$ as $k'=4k$. As a consequence the cosmological constant is
${8\over k'}$ in eqn. (14) whereas it is ${2\over k}$ in the action of
Ref. [9]. The relation between the three dimensional black string
solution and the charged black hole solution can be envisaged from
the point of view of duality transformations.
Since the backgrounds are independent of the coordinate $x$, there is
$x$ translation symmetry. Thus one can apply the duality
transformations on (12,13) and get (19,20). The equivalence
between momentum and axionic charge was discussed in Ref. [18] by
using the arguments of duality transformation. In (21) if one does not
assume $x$ to be
compact one can think the $g_{xt}$ term in eqn. (24a) as momentum
along $x$ direction.
The $O(2,2)$ transformation we have used in
this case is equivalent to duality transformation (10). Although we
can arrive at solution (25) from solution (12) through an $O(d,d)$ or
duality trnsformation but the underlying conformal field theories in
two cases are very different because in eqn. (12) $x$ is non-compact
whereas in eqn. (24) the $x$ direction was taken to be compact to
arrive at solution (25) (Ref. [17]).The space-time
geometry and other properties of these solutions are extensively
discussed in Refs. [5,9].

Now we consider the construction of monopole background in the string
effective action
discussed by Banks $et al.$ [15]. The string effective action is a
generalisation of the Kaluza-Klein theory considered by Sorkin, Gross
and Perry [6]. The string effective action could arise from the
following scenario for a closed string. We can have a configuration
as envisaged by Gaspirini, Maharana and Veneziano [13], that 21 space
dimensions of the bosonic string are flat and of the remaining five
dimensions, one corresponding to $x^{22}$, is compactified  with
radius $R_0$. The five dimension theory (before compactification) is
endowed with graviton, antisymmetric tensor field and a dilaton
satisfying the equations of motion required to satisfy conformal
invarance. When the coordinate $x^{22}$ is compactified the massless
spectrum consists of a spacetime graviton ($g_{\mu\nu}$),
antisymmetric field ($b_{\mu\nu}$), and two gauge fields $A_{\mu}$
and $B_{\mu}$ coming from five dimensional metric and antisymmetric
tensor field. Moreover, there are two scalar fields $\Phi$ and
$g_{55}=R$. The equations of motion for the background fields are
derived from the action,

$$ S=\int d^{5}x \sqrt{detG}
e^{-\phi}\big [R^{(5)}+G^{\mu\nu}\partial_{\mu}\phi\partial_{\nu}\phi-{1\over
12}H^2 \big ],\eqno(26)$$
\noindent where $R^{(5)}$ is the five dimensional scalar curvature.
The indices should run over five dimensions. The specific background
fields which are cosistent with the vanishing $\beta$-function
conditions are given by

$$ds^2 = -dt^2 + R^2(dx^5 + A_{\phi}d\phi)^2 + {1\over {R^2}}(dr^2 +
r^2d\Omega^2) , \eqno (27)$$
\noindent where $A_{\phi} = {R_0}\sin^2({\theta\over 2})$ and $R^2 = (1 +
{{R_0}\over {2r}})^{-1}$. The dilaton  $\Phi = \Phi_{0} = const.$, and
the antisymmetric tensor field $B = 0$. The direction $x^5$ is compact.
So $A_{\phi}$ here is like a gauge field.
This solution was obtained by solving $R_{MN} = 0$, where $M$ and $N$
run over all the five coordinates, so in five dimension it is a Ricci
flat solution. One can verify that this field configuration satisfies
the equations of motion following from the effective action (26).
The above metric is independent of three coordinates: $t$,
$x$, and $\phi$. So one has the $O(3,3)$ symmetry. $O(3,3)$ rotation can be
performed on $t-\phi-x^5$ block and new solution can be generated.
Again we use the same $ \Omega$ (defined earlier in
(15)). The $O(2)\times O(2)$ is a subgroup of $O(3,3)$.
We want to perform this transformation on the $\phi-x^5$ block of metric.
The $\phi-x^5$ block of the metric (27) reads,

$$\clg = \Bigg ( \matrix{{{r^2\sin^2\theta {R^{-2}}} + A_{\phi}^{2}R^2
} & {A_{\phi}R^2}\cr
{A_{\phi}R^2} & {R^2} \cr}
\Bigg ), \qquad {\cal B} = 0. \eqno (28)$$

After the action of $\Omega$ on this the new generated metric and
antisymetric field tensor are

$$\clg' = \Bigg ( \matrix {{{r^2\sin^2\theta  {R^{-2}}}} & {0}\cr
{0} & {R^{-2}} \cr}
\Bigg ),\qquad {\cal B}' = \Bigg ( \matrix {{0} & {A_{\phi}}\cr
{-A_{\phi}} & {0} \cr}
\Bigg ), \eqno (29a)$$
\noindent and new dilaton is,
$$\Phi = \Phi_{0} + \ln ({1\over {R^2}}). \eqno (29b)$$
So the full metric and torsion field can be written as,

$$ds^2 = -dt^2 + {1\over R^2}{dx^5}^2 + {1\over R^2}(dr^2 +
r^2d\Omega^2),\qquad  B_{\phi 5} = A_{\phi} \eqno (30)$$
\noindent with $R^2  = (1 + {R_0\over 2r})^{-1}$.
This is exactly the dual solution obtained by Banks $et al.$ [15]
using the symmetry properties of the effective action. It is
interesting to note that this solution has non trivial antisymmetric
field tensor component as well as non trivial dilaton, whereas in the
original solution both this fields were trivial. This solution
is similar to the original solution of first example. We observe that
in both the cases the particular $\Omega (O(2)\times O(2))$
interchanges gauge fields and antisymmetric field tensor. Again
solution (30) can be obtained from the Sorkin, Gross and Perry
solution (27) using the duality transformations (10). For this one
uses the translation symmetry of $x^5$ and assumes $x\equiv x^5$
in (10). The dual solution is also a monopole solution  and
this time it is magnetic. Its properties etc are discussed in [15].

\noindent {\bf IV. Summary and Conclusions}

The $O(d,d)$ transformations transform a given string background
geometry to another (in general inequivalent) geometry. Here we have
presented two examples where appropiate $O(d,d)$ transformation
corresponds to duality transformation eqn. (10). The $O(2,2)$
transformations employed  by us interchange the gauge field and the
antisymmetric tensor field in both the cases. In the first example we
show that the axionic charge (in a given background configuration)
can be transformed in another background with electric charge in a
lower dimension.We are also able to relate two different backgrounds (12) and
(25), which were obtained very
differently, via $O(2,2)$ (equivalently duality) transformation.

\bigskip
\bigskip
\smallskip

\noindent {\bf Acknowledgements:} One of us (JM) would like to
acknowledge useful discussions with John Schwarz and he would like to
thank the High Energy Physics group, especially John Schwarz, for
warm hospitality at Caltech.

\vfil
\eject

 \noindent {\bf References:}

\item {[1]} C. G. Callan, D. Friedan, E. J. Martinec and M. J. Perry,
{\it Nucl. Phys.} {\bf B262} (1985) 593; A. Sen, {\it Phys. Rev. Lett.}
{\bf 55} (1985) 1846.

\item {[2]} A. A. Tseylin and C. Vafa, {\it Nucl. Phys.} {\bf B372} (1992) 443;
A. A. Tseytlin Cambridge Univ. preprint {\it DAMTP}-37-1991.

\item {[3]} G. W. Gibbons and K. Maeda, {\it Nucl. Phys.} {\bf B298}
(1988) 741;
C. G. Callan, R. C. Myers and M. J. Perry, {\it Nucl. Phys.} {\bf B311}
(1988/89) 673; H. J. Vega and N. Sanchez, {\it Nucl. Phys.} {\bf B309}
(1988) 552; {\it Nucl. Phys.} {\bf B309} (1988) 557;
G. Mandal, A. M. Sengupta and S. Wadia, {\it Mod. Phys. Lett.}
{\bf A6} (1991) 1685; D. Garfinkle, G. T. Horowitz and A. Strominger,
{\it Phys. Rev.} {\bf D43} (1991) 3140;
S. P. de Alwis and J. Lykken, {\it Phys. Lett.} {\bf 269B} (1991) 264;
A. Tseytlin, Johns Hopkins preprint {\it JHU-TIPAC}-91009 (1991); A. Giveon,
LBL preprint {\it LBL}-30671 (1991).

\item {[4]} S. B. Giddings and A. Strominger, {\it Phys. Rev. Lett.}
{\bf 67} (1991) 2930;
G. T. Horowitz and A. Strominger, {\it Nucl. Phys.} {\bf B360}
(1991) 197; M. J. Duff and J. X. Lu, {\it CTP-TAMU}-81/90 (1990);
A. Sen, {\it Phys. Lett.} {\bf 274B} (1991) 34.

\item {[5]} J. H. Horne and G. T. Horowitz,  {\it Nucl. Phys.} {\bf B368}
(1992) 444.

\item {[6]} R. Sorkin, {\it Phys. Rev. Lett.} {\bf 51} (1983) 87; D. J. Gross
and M. J. Perry, {\it Nucl. Phys.} {\bf B226}
(1983) 29.

\item {[7]} J. A. Harvey and J. Liu, {\it Phys. Lett.} {\bf 268B}
(1991) 40; A. Shapere, S. Trivedi and F. Wilczek, Institute of
Advanced Study, Princeton preprint {\it IASSNS-HEP}-91/33 (1991).

\item {[8]} E. Witten, {\it Phys. Rev.} {\bf D44} (1991) 314;
R. Dijgraaf, E. Verlinde and H. Verlinde, Institute for Advanced Study,
Princeton preprint {\it IASSNS-HEP}-91/22 (1991);
E. B. Kiritsis, {\it Mod. Phys. Lett.} {\bf A6} (1991) 2871;
I. Bars, University of South California preprint, {\it USC}-91/{\it HEP}-B3
(1991); {\it USC}-91/{\it HEP}-B4 (1991);
E. Martinec and S. Shatasvili, {\it Nucl. Phys.} {\bf B368}
(1992) 338;
M. Bershadsky and D. Kutasov, {\it PUPT}-1261, {\it HUTP}-91/A024;
P. Horava, Chicago preprint {\it EFI}-91-57 (1991); S. Nojiri,
{\it FERMILAB-PUB}-91/230-{\it T}, KEK preprint 91-99, {\it KEK-TH}-388.

\item {[9]} N. Ishibashi, M. Li and A. R. Steif, {\it Phys. Rev. Lett.}
{\bf 67} (1991) 3336.

\item {[10]} K. Kikkawa and M. Yamasaki, {\it Phys. Lett.} {\bf 149B}
(1984) 357; N. Sakai and I. Senda, {\it Prog. Theor. Phys.} {\bf 75}
(1986) 692; V. Nair, A. Shapere, A. Stominger and F. Wilczek,
{\it Nucl. Phys.} {\bf B287} (1987) 402; A. Giveon, E. Rabinovici
and G. Veneziano, {\it Nucl. Phys.} {\bf B322} (1989) 167; For a
recent review see $``$Space-time duality in string theory", J. H.
Schwarz in Elementary Particles and the Universe, Essays in Honour of
Murray Gell-Mann, ed. J. H.
Schwarz, Cambridge University Press 1991 and references therein, and
Caltech Preprint, {\it CALT}-68-1740, May 1991.

\item {[11]} A. A. Tseytlin, $``$Space-time duality, dilaton and
string cosmology", Proc. of the First International A. D. Sakharov
Conference on Physics, Moscow 27-30 May 1991, ed. L. V. Keldysh
et al. Nova Science Publ., Commack, N. Y., 1991.

\item {[12]} K. A. Meissner and G. Veneziano, {\it Phys. Lett.} {\bf 267B}
(1991) 33; G. Veneziano, {\it Phys. Lett.} {\bf 265B} (1991) 287.
 K. A. Meissner and G. Veneziano, {\it Mod. Phys. Lett.} {\bf A6}
(1991) 3397.

\item {[13]} M. Gasperini, J. Maharana and G. Veneziano,
{\it Phys. Lett.} {\bf 272B} (1991) 167;
M. Gasperini and G.Veneziano, {\it Phys. Lett.} {\bf 277B} (1991) 256.

\item {[14]} A. Sen, {\it Phys. Lett.} {\bf 271B} (1991) 295;
A. Sen, {\it Phys. Lett.} {\bf 274B} (1991) 34;
S. F. Hassan and A. Sen, Tata preprint, {\it TIFR/TH/}91-40,
(1991) (to appear in {\it Nucl. Phys. B});
S. P. Khastgir and A. Kumar, {\it Mod. Phys. Lett.} {\bf A6}
(1991) 3365; S. K. Kar, S. P. Khastgir and A. Kumar, Institute of
Physics, Bhubaneswar preprint, {\it IP/BBSR}/91-51 (1991) (to appear
in {\it Mod. Phys. Lett. A}); A. Sen, Tata preprint, {\it TIFR/TH/}92-20,
(1992).

\item {[15]} T. Banks, M. Dine, H. Dijkstra and W. Fischler,
{\it Phys. Lett.} {\bf 212B} (1988) 45.

\item {[16]} T. Busher, {\it Phys. Lett.} {\bf 194B} (1987) 59;
{\it Phys. Lett.} {\bf 201B} (1988) 466;
{\it Phys. Lett.} {\bf 159B} (1985) 127.

\item {[17]} M. Rocek and E. Verlinde, {\it Nucl. Phys.}
{\bf B373} (1992) 630.

\item {[18]}J. H. Horne, G. T. Horowitz and A. R. Steif, {\it Phys. Rev. Lett.}
{\bf 68} (1991) 568.

\end